\documentclass[aps,pra,reprint,amsmath,amssymb,showkeys,notitlepage,superscriptaddress]{revtex4-2}
\usepackage{color}
\usepackage{braket}
\usepackage{graphicx} 
\usepackage{txfonts}  
\usepackage{epstopdf}
\usepackage{appendix}
\usepackage{bigints}
\usepackage{stackengine}
\usepackage{xcolor}
\usepackage{mathtools}
\usepackage{bbold}
\usepackage{braket}
\begin{document}
\title{Heisenberg treatment of multiphoton pulses in {\color{black} waveguide QED with time-delayed feedback}}

\author{Kisa Barkemeyer}
\thanks{k.barkemeyer@tu-berlin.de}
\affiliation{Institut f\"ur Theoretische Physik, Technische Universit\"at Berlin, D-10623 Berlin, Germany}
\author{Andreas Knorr}
\affiliation{Institut f\"ur Theoretische Physik, Technische Universit\"at Berlin, D-10623 Berlin, Germany}
\author{Alexander Carmele}
\affiliation{Institut f\"ur Theoretische Physik, Technische Universit\"at Berlin, D-10623 Berlin, Germany}

\begin{abstract}
{\color{black}
The dynamics of waveguide-QED systems involving coherent time-delayed feedback
give rise to a hierarchy of multi-time correlations within the Heisenberg picture due to the induced non-Markovianity.
We propose to perform a projection onto a complete set of states in the Hilbert space to decompose the multi-time correlations into single-time matrix elements.
To illustrate the procedure, we consider the paradigmatic example of a two-level system that couples to a semi-infinite waveguide and interacts with quantum light pulses.
Our approach complements the range of available methods as it 
allows calculating the dynamics under the inclusion of additional dissipation channels in a numerically exact and efficient manner for multiphoton pulses of arbitrary shape where memory requirements are known in advance.
}
\end{abstract}

\maketitle

\section{Introduction}

Quantum networks, in which quantum information is transferred between flying and stationary qubits, form the basis for a plethora of applications in quantum computation and communication \cite{Cirac1997a,DiVincenzo2000,Zoller2005,Kimble2008}. 
In this context, waveguide quantum electrodynamics (WQED) systems are of particular interest as model systems for the interaction of quantum emitters with the electromagnetic field inside a one-dimensional geometry \cite{Chang2007,OBrien2009,Northup2014,Roy2017,Vermersch2017}. 
In large-scale quantum networks where the spatial separations are non-negligible in comparison to the wavelength of the light, the information backflow from the environment into the system has to be accounted for and renders the dynamics non-Markovian \cite{Breuer2016,DeVega2017,Fang2018,Sinha2020}.
These delay times can play a constructive role since they allow the control of the system by means of coherent time-delayed feedback where the quantum state itself steers the dynamics and no measurement is necessary so that quantum coherence is preserved \cite{Wiseman1994,Lloyd2000,Carmele2013,Hein2014,Grimsmo2015,Hoi2015a,Nemet2016,Pichler2017,Carmele2020,Katsch2020}.

Non-Markovian feedback in WQED systems generates strongly entangled system-reservoir states which are intractable with perturbative and master equation methods typically used in the Markovian case \cite{Carmichael1993,Shi2015,Dinc2019,Kiilerich2019}. Besides the brute force integration of the full Schr\"odinger equation \cite{Hein2014,Lu2017}, feasible only for very few excitations and short-time dynamics, in recent years, two main schemes have been proposed to deal with such systems:
The first one, an optimized Schr\"odinger equation method, relies on the evaluation of the quantum stochastic Schr\"odinger equation within the matrix product state framework \cite{Pichler2016,Guimond2017,ArranzRegidor2021}. Here, the dynamics are cast into a time-discrete basis and justified truncations of the Hilbert space enable the efficient handling of the entanglement. {\color{black} This way, the numerically exact method allows for long-time simulations and fast convergence times. The extension to non-hermitian dynamics is, however, involved and numerically expensive \cite{Verstraete2004,Zwolak2004,Daley2009,Kaestle2021}.} Furthermore, as the number of necessary Schmidt values is not known without convergence analysis, the required memory cannot be determined beforehand. The second method is based on the von Neumann equation and exploits a series of cascaded quantum systems to account for the system being driven by its own past \cite{Grimsmo2015,Whalen2017}. This insightful approach is highly expensive memory-wise as it scales with the exponentially growing dimension of the corresponding Liouvillian and, thus, is limited to the study of small systems and short times.
Moreover, factorization schemes have not been implemented so far due to the tedious construction of the corresponding Liouvillian. 
%Recently, another approach for studying coherent time-delayed feedback has been proposed that relies on the treatment of the space-discretized waveguide using quantum trajectory simulations \cite{ArranzRegidor2021,Crowder2021}.
%
Complementing existing methods, we present a numerically exact Heisenberg method using a time-discrete basis where the memory requirements are known in advance.
The approach allows for the inclusion of dissipation and many-particle dynamics as well as approximation methods without the cost of losing exactness in the coherent quantum feedback dynamics. 

When treating non-Markovian WQED systems in the Heisenberg picture, the time non-locality usually introduces a hierarchy of multi-time correlations \cite{DeVega2017}. 
For the efficient calculation of the dynamics of arbitrary WQED systems excited via quantum pulses and subjected to coherent time-delayed feedback, we propose to unravel the emerging multi-time correlations via the insertion of a Hilbert space unity between the operators with different time arguments. This corresponds to a projection onto a complete set of states in the Hilbert space. In this way, we decompose the multi-time correlations and obtain a closed system of differential equations for single-time correlations. 

The paper is structured as follows: After this introduction, in Sec.~II, we present the proposed Heisenberg method using the example of a two-level system (TLS) coupled to a semi-infinite waveguide.  We benchmark the dynamics that can be obtained this way and compare it to related approaches for the treatment of WQED setups with time-delayed feedback. In Sec.~III, we use our approach to study the impact of multiphoton pulses of variable shape as well as the effect of a phenomenological pure dephasing as an exemplary additional dissipation channel. Finally, we conclude in Sec.~IV.

\section{Method}

We introduce our Heisenberg approach for the calculation of the exact feedback dynamics in WQED systems using the fundamental example of a TLS coupled to a semi-infinite one-dimensional waveguide as depicted in Fig.~\ref{fig:Schema}\,a \cite{Cook1987,Dorner2002,Tufarelli2013}. The closed end of the waveguide at the distance $d$ from the TLS functions as a mirror. It provides feedback at the delay time $\tau = 2d/c$ where $c$ is the speed of light in the waveguide. The quantized treatment of the light field allows modeling the excitation of the TLS through the waveguide via a quantum pulse of temporal shape $f(t)$. 
In the dipole as well as the rotating wave approximation, the Hamiltonian describing the system after a transformation into the rotating frame defined by its non-interacting part takes the form \cite{Walls2008}
\begin{equation}
    \mathcal{H}' = \hbar \int d\omega  g(\omega) \left( e^{i \left(\omega -\omega_0\right)t} r_\omega^\dagger \sigma_-  + \text{H.c.}\right).
    \label{eq:H_int}
\end{equation}
Here, the operator $\sigma_-$ ($\sigma_+$) is the lowering (raising) operator of the TLS with transition frequency $\omega_0$, $\sigma_+ = \sigma_-^\dagger$. Under the assumption that only a single excitation in the TLS is possible, the commutator of these operators is $\left[\sigma_-,\sigma_+ \right] = \mathbb{1}-2\sigma_+\sigma_-$.
The annihilation (creation) of a photon with frequency $\omega$ in the reservoir is described by the bosonic operator $r_\omega^{(\dagger)}$.
The interaction between the TLS and the reservoir is quantified by the coupling strength $g(\omega)$ which is, in general, frequency dependent.
In particular, the feedback mechanism in our system imposes a boundary condition and results in a sinusoidal frequency dependence of the coupling strength, $g(\omega) = g_0 \sin(\omega \tau/2)$ \cite{Cook1987,Dorner2002,Carmele2013,Tufarelli2013}.

\subsection{Heisenberg treatment}

To model the system dynamics, we use the Hamiltonian in Eq.~\eqref{eq:H_int} to derive differential equations for the time-dependent operators $\sigma_-(t)$ and $r_\omega(t)$ via the Heisenberg equation of motion. 
The explicit reservoir dynamics can be eliminated by integrating out the reservoir operator $r_\omega(t)$,
\begin{equation}
    r_\omega(t) = r_\omega(0) - i g(\omega) \int_0^t dt' e^{i(\omega-\omega_0)t'} \sigma_-(t').
\end{equation}
Thus, the occupation of the reservoir modes is governed by the initial occupation of the reservoir which encodes the pulse driving the emitter as well as by the interaction of the reservoir with the TLS.

The initial occupation of the reservoir is transformed into the time domain via
\begin{equation}
    r_t = \frac{1}{\sqrt{2\pi}}\int d \omega r_\omega(0) e^{-i\left(\omega-\omega_0\right)t},
    \label{eq:noise_op}
\end{equation}
which can be interpreted as an input operator in the context of the input-output theory \cite{Gardiner}. While the operator $r_\omega(0)$ annihilates a photon of frequency $\omega$, the operator $r_t$ as its Fourier transform annihilates a photon at time $t$.
%<
This way, for the TLS operator $\sigma_-$ the delay differential equation
\begin{multline}
    \frac{\text{d}}{\text{d}t} \sigma_-(t) = - \Gamma \sigma_-(t)  - \sqrt{\Gamma}\left[ \mathbb{1} - 2 \sigma_+(t)\sigma_-(t) \right] r_{t,\tau} \\
    +\Gamma e^{i \omega_0 \tau}\left[\sigma_-(t-\tau) - 2 \sigma_+(t) \sigma_-(t) \sigma_-(t-\tau) \right] \Theta(t-\tau) \label{eq:P_op}
\end{multline}
with the decay rate $\Gamma \equiv \pi g_0^2/2$ and the delayed input operator $r_{t,\tau} \equiv r_{t-\frac{\tau}{2}}e^{i \omega_0 \frac{\tau}{2}}-r_{t+\frac{\tau}{2}}e^{-i\omega_0 \frac{\tau}{2}}$ can be determined. Eq.~\eqref{eq:P_op} implies that a photon emitted from the TLS towards the mirror is reflected and interacts with the TLS again after the delay time $\tau$. Thus, for times $t \geq \tau$, feedback effects come into play since the feedback signal interferes with the emission from the TLS where the impact of the feedback depends on the feedback phase $\omega_0 \tau$.

\begin{figure}
    \centering
    \includegraphics[width=8.0cm]{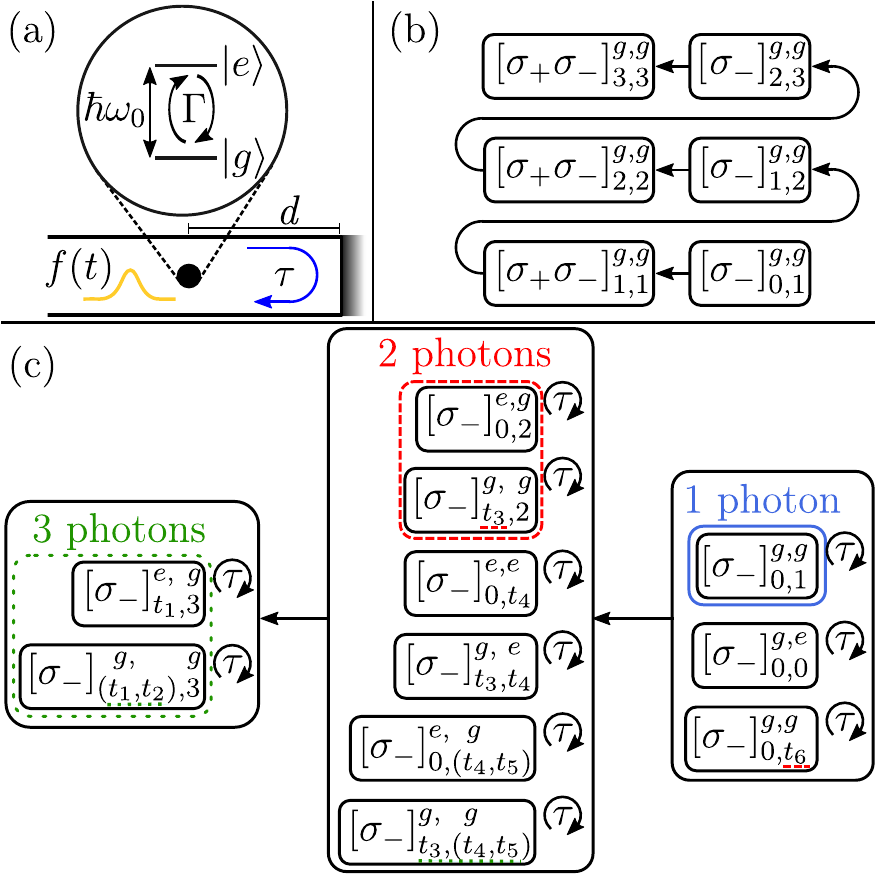}
    \caption{(Color online) {\color{black}
    TLS subjected to coherent time-delayed feedback and the resulting hierarchy of matrix elements. (a) Illustration of the setup which consists of a TLS with transition energy $\hbar \omega_0$ between ground state $\ket{\text{g}}$ and excited state $\ket{\text{e}}$ coupling with decay rate $\Gamma$ to a semi-infinite waveguide. The closed end of the waveguide at distance $d$ from the TLS provides feedback at the delay time $\tau$ and the TLS can be excited through the waveguide via a quantum pulse of temporal shape $f(t)$.  (b) Scheme of the contributing matrix elements for the dynamics without feedback for an excitation with up to three photons. The abbreviation $\left[ A \right]_{m,n}^{i,j}$, $A  \in \{ \sigma_+\sigma_-,\sigma_-\}$, encodes the matrix element $\bra{i,m}A(t)\ket{j,n}$. The $n$-photon dynamics can be obtained recursively from the case of $n-1$ photons.  (c) Scheme of the contributing matrix elements for the dynamics with feedback. When decomposing the expectation value of the TLS population, the matrix elements in the colored boxes have to be calculated and saved for the duration of one feedback interval $\tau$.
    The elements $\left[ A \right]_{m,n}^{i,j}$ can be calculated recursively where, however, additionally the matrix elements outside the colored boxes couple to the dynamics.
    The underlined times indicate the scaling of the algorithm for the corresponding number of photons since they have to be integrated over and determine the required computational resources.}}
    \label{fig:Schema}
\end{figure}

The observable of interest is the population of the excited state of the TLS, $\sigma_+(t)\sigma_-(t)$, which we will refer to as the emitter population in what follows with the corresponding expectation value $\bra{\psi}\sigma_+(t)\sigma_-(t)\ket{\psi}$. The initial state $\ket{\psi}$ is the combined state of the TLS and the reservoir. For demonstration purposes, we assume a separable pure initial state of the form $\ket{\psi} = \ket{j} \otimes \ket{n} \equiv \ket{j,n}$ which denotes the TLS either in the ground ($j = g$) or the excited state ($j = e$) and $n$ photons in the reservoir, $n \in \mathbb{N}$. Note, however, that our method is not limited to this class of states but is also applicable in the case of entangled and mixed states.
The $n$-photon state $\ket{j,n}$ can be obtained from the vacuum state of the reservoir via \cite{Loudon,Raymer2020} 
\begin{equation}
    \ket{j,n} = \frac{1}{\sqrt{n!}}\left(a_f^\dagger\right)^n\ket{j,0}, \quad a_f^\dagger = \int dt f(t) r^\dagger_t.
    \label{eq:wavepacket_creation}
\end{equation}
Here, $a_f^\dagger$ is the creation operator of a photon in the temporal mode defined by a
wave packet with normalized pulse shape $f(t)$ satisfying
\begin{equation}
    \int dt \left|f(t) \right|^2 = 1
\end{equation}
so that the creation of a single excitation via $a_f^\dagger$ in Eq.~\eqref{eq:wavepacket_creation} is ensured.
When calculating the expectation value of the TLS population, we have to solve the differential equation
\begin{align}
    &\frac{\text{d}}{\text{d}t} \bra{j,n}\sigma_+(t)\sigma_-(t) \ket{j,n} = - 2 \Gamma \bra{j,n}\sigma_+(t)\sigma_-(t) \ket{j,n} \notag \\
    &- \sqrt{n \Gamma}  \left[f^*_{\tau}(t) \bra{j,n-1} \sigma_{-}(t)\ket{j,n} +  \text{H.c.} \right]\notag \\
    &+ \Gamma\left[e^{-i\omega_0 \tau} \bra{j,n}  \sigma_{+}(t-\tau)\sigma_{-}(t)\ket{j,n} +  \text{H.c.}\right] \Theta(t-\tau) \label{eq:E_mat}
\end{align}
with $f_\tau(t) = f(t-\frac{\tau}{2})e^{i\omega_0\frac{\tau}{2}} - f(t+\frac{\tau}{2})e^{-i\omega_0\frac{\tau}{2}}$. In Eq.~\eqref{eq:E_mat}, the expectation value of the TLS population couples to the single-time matrix element of the TLS operator $\bra{j,n-1}\sigma_-(t)\ket{j,n}$ which, in turn, obeys
\begin{align}
    \frac{\text{d}}{\text{d}t} &\bra{j,n-1}\sigma_{-}(t)\ket{j,n} = - \Gamma \bra{j,n-1}\sigma_{-}(t)\ket{j,n}   \notag \\
    &- \sqrt{n \Gamma} f_{\tau}(t)\left[ 1 - 2 \bra{j,n-1}\sigma_+(t)\sigma_-(t) \ket{j,n-1}\right] \notag \\
    &+ \Gamma e^{i \omega_0 \tau}\left[\bra{j,n-1}\sigma_{-}(t-\tau)\ket{j,n} \notag \right. \\
    &\quad\left. - 2 \bra{j,n-1}\sigma_+(t)\sigma_-(t) \sigma_{-}(t-\tau)\ket{j,n} \right] \Theta(t-\tau). \label{eq:P_mat}
\end{align}

Before the feedback mechanism sets in, that is, for $t<\tau$ as built-in by $\Theta(t-\tau)$, we obtain time-local and, thus, essentially Markovian dynamics. In this case, the dynamics for $n$ photons in the reservoir initially can be calculated by recursively inserting the results for $n-1$ photons \cite{Wang2012,Barkemeyer2021}. The equations for the contributing matrix elements are given explicitly in Appendix~A. For a TLS initially in the ground state and a three-photon pulse in the reservoir, they are exemplarily sketched in Fig.~\ref{fig:Schema}\,b  where the abbreviation $\left[ A \right]_{m,n}^{i,j}$ encodes the matrix element $\bra{i,m}A(t)\ket{j,n}$.

For later times, that is, for $t\geq \tau$, when the feedback mechanism influences the dynamics, however, the time-delayed terms in Eqs.~\eqref{eq:E_mat} and \eqref{eq:P_mat} become relevant, and the single-time correlations couple to two-time correlations. In general, these two-time correlations couple to three-time correlations and so on. To deal with this hierarchical structure, we unravel the multi-time correlations via the insertion of a unity operator, that is, a projector onto a complete set of states in the Hilbert space, between the operators with different time arguments.
In the time basis, this unity operator takes the form
\begin{multline}
    \mathbb{1} = \big\{\ket{g}\bra{g} + \ket{e}\bra{e} \big\} \otimes \\ 
    \Bigg\{ \ket{0}\bra{0} + \int dt' \ket{t'}\bra{t'} + \frac{1}{2} \int dt' \int dt'' \ket{t',t''}\bra{t',t''} + \dots\Bigg\}.
    \label{eq:unity}
\end{multline}
The reservoir state $\ket{t'}$ denotes a single photon in the reservoir created at time $t'$, $\ket{t'} = r_{t'}^\dagger \ket{0}$. Here, as in the following, we generally consider all times $t' \in \mathbb{R}$. 
The state $\ket{t',t''}$ refers to two photons in the reservoir, one created at time $t'$, the other at time $t''$, $\ket{t',t''} = r_{t'}^\dagger r_{t''}^\dagger \ket{0}$, etcetera. The photons are indistinguishable so that $\ket{t',t''} = \ket{t'',t'}$ and the factor $1/2$ (in general $1/(n!)$ for $n$ photons in the reservoir) is required to ensure normalization.
Inserting Eq.~\eqref{eq:unity} between operators at different times in Eq.~\eqref{eq:P_mat}, we avoid the explicit calculation of multi-time correlations with time-ordering and, this way, shift the problem to the calculation of the matrix elements of single-time Heisenberg operators. Furthermore, we can insert the unity operator between the TLS operators $\sigma_\pm$ at the same time to only have to deal with matrix elements of the TLS operator $\sigma_-$.

{\color{black}
It depends on the number of excitations in the system which elements of the unity operator in Eq.~\eqref{eq:unity} effectively contribute to the dynamics. 
For simplicity, in what follows, we focus on the case where the TLS is initially in the ground state so that all excitations are in the reservoir. 
However, the results can be extended to the case of an initially excited TLS with increased numerical complexity.}
We start with the decomposition of the expectation value of the TLS population, that is, of the single-time matrix element
\begin{equation}
\bra{g,n}\sigma_+(t)\sigma_-(t) \ket{g,n} = \bra{g,n}\sigma_+(t)\mathbb{1}\sigma_-(t) \ket{g,n}.
\label{eq:Unity_exp}
\end{equation}
Here, we find that only matrix elements of the form $\bra{\psi_{n-1}} \sigma_-(t)\ket{g,n}$
contribute to the dynamics where $\ket{\psi_{n-1}}$ is a shorthand notation for a state comprising $n-1$ excitations.
The state $\ket{\psi_{n-1}}$ can describe either an excitation in the TLS and $n-2$ excitations in the reservoir or zero excitations in the TLS and $n-1$ excitations in the reservoir.
As a consequence, only projectors onto such states $\ket{\psi_{n-1}}$ have to be taken into account in the unity operator.
Subsequently, we use this result to decompose all matrix elements so that we only have to deal with single-time correlations of the TLS operator $\sigma_-$.

As in the case without feedback, the results with feedback for $n$ photons in the reservoir initially can be obtained recursively by using the results for $n-1$ photons in the reservoir. However, to account for the implemented feedback mechanism, further matrix elements have to be evaluated as sketched in Fig.~\ref{fig:Schema}\,c for up to three photons in the reservoir. 
Correspondingly, we outline the approach, starting with a single photon and progressing to two and three photons in the pulse. Higher excitation numbers can be treated analogously. The explicit equations for the relevant matrix elements are deferred to Appendix~B.

In the single-excitation limit, only the projector onto the state $\ket{g,0}$ describing the TLS in the ground state and zero photons in the reservoir contributes when inserting the unity operator into the TLS population.
Consequently, we can decompose the TLS population according to
\begin{multline}
\bra{g,1}\sigma_+(t)\sigma_-(t) \ket{g,1} \\ = \bra{g,1}\sigma_+(t)\ket{g,0}\bra{g,0}\sigma_-(t) \ket{g,1}
\end{multline}
(Fig.~\ref{fig:Schema}\,c, black, solid box).
The computational resources required for the calculation of the dynamics in the single-excitation case grow linearly with the number of time steps $N$, $\mathcal{O}(N)$.

With two excitations in the system, the elements in the unity operator describing a single excitation in the system contribute. This excitation can be either in the TLS or in the reservoir so that
\begin{multline}
    \bra{g,2}\sigma_+(t) \sigma_-(t) \ket{g,2} = \bra{g,2}\sigma_+(t)\ket{e,0}\bra{e,0} \sigma_-(t)\ket{g,2} \\
    + \int dt' \bra{g,2} \sigma_+(t) \ket{g,t'}\bra{g,t'} \sigma_-(t) \ket{g,2}
    \label{eq:2ex}
\end{multline}
(Fig.~\ref{fig:Schema}\,c, red, dashed box). Note that times $t'$ for which it holds that $f(t') = 0$ cannot be omitted in the integral since the corresponding matrix elements can still yield a noise contribution although there is no pulse contribution. In particular, negative times $t'<0$ have to be considered which account for the vacuum noise initially in the feedback channel.
As indicated by the red dashed underlined times in Fig.~\ref{fig:Schema}\,c, we have to perform two nested integrations in each time step. Thus, the computational resources required when considering two excitations in the system grow with the number of time steps $N$ to the power of three, $\mathcal{O}\left(N^3\right)$. 

If there are three excitations in the system, all states describing two excitations in the system in the unity operator contribute to the dynamics and we can decompose the TLS population according to
\begin{multline}
     \hspace{-1em}\bra{g,3}\sigma_+(t) \sigma_-(t) \ket{g,3} = \!\!
     \int \!\! dt' \bra{g,3}\sigma_+(t)\ket{e,t'}\bra{e,t'} \sigma_-(t)\ket{g,3} \\
    + \frac12 \int dt'\hspace{-0.4em}\int dt'' \bra{g,3} \sigma_+(t) \ket{g,t',t''}\bra{g,t',t''} \sigma_-(t) \ket{g,3}
    \label{eq:3ex}
\end{multline}
(Fig.~\ref{fig:Schema}\,c, green, dotted box). A feature that becomes important if at least three excitations in the system are considered, is the indistinguishability of the photons in the states the unity operator projects onto. 
For three excitations in the system, the calculation of the dynamics requires computational resources growing with the number of time steps $N$ to the power of six, $\mathcal{O}\left(N^6\right)$, as indicated in Fig.~\ref{fig:Schema}\,c by the green dotted underlined times. {\color{black} Generalizing to the case of $n>1$ excitations in the system, the number of matrix elements that have to be calculated grows with the number of time steps $N$ to the power of $3(n-1)$, $\mathcal{O}\left(N^{3(n-1)}\right)$.}

\subsection{Benchmark}
\label{sec:Benchmark}

\begin{figure}
    \centering
    \includegraphics[width=8.6cm]{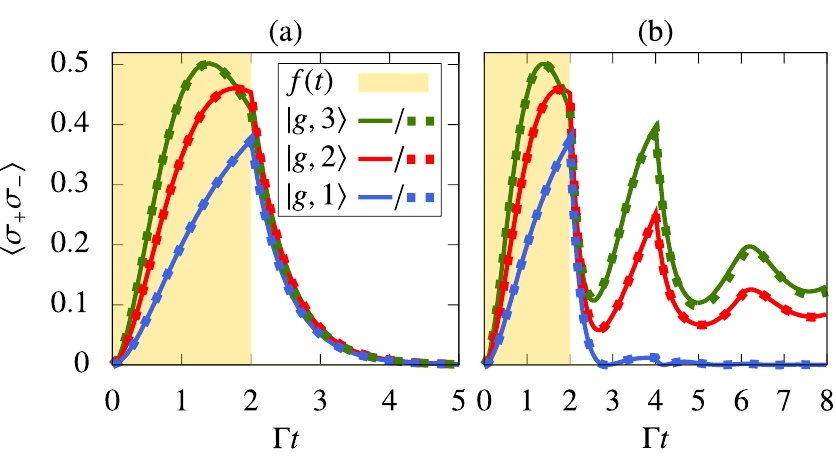}
    \caption{(Color online) Benchmark for the dynamics of the TLS population in the initial state $\ket{g,n}$ with a rectangular pulse of duration $\Gamma t_D = 2$ (yellow, shaded area) containing $n$ photons, $n \in \{1,2,3\}$. (a) Benchmark for the decomposition of the TLS population without feedback where the expectation value is either directly calculated (solid lines) or decomposed via the insertion of a unity operator (dashed lines). (b) Benchmark of the feedback dynamics for feedback at $\Gamma \tau = 2$ with feedback phase $\omega_0\tau = 2\pi m$, $m \in \mathbb{N}$, calculated using the matrix product state framework (solid lines) or the Heisenberg method (dashed lines).}
    \label{fig:Benchmark}
\end{figure}

To verify the validity of our approach, we benchmark the dynamics without as well as with feedback.
We use the dynamics we obtain before the feedback sets in, $t < \tau$, from the direct calculation of the expectation value of the TLS population as given in Appendix~A to ensure the validity of the decomposition of the TLS population via the insertion of a unity operator. In this regime, the dynamics are essentially Markovian. The dynamics for a TLS excited by a rectangular pulse of duration $\Gamma t_D = 2$ (yellow, shaded area) containing up to three photons are presented in Fig.~\ref{fig:Benchmark}\,a. We see that the results without (solid lines) and with the decomposition of the expectation value via the insertion of a unity operator (dashed lines) coincide perfectly indicating the correctness of the decomposition.

In the next step, we turn to the non-Markovian regime, that is, we include times $t\geq\tau$ where the feedback mechanism influences the dynamics. In Fig.~\ref{fig:Benchmark}\,b, the dynamics for a TLS subjected to feedback at delay time $\Gamma \tau = 2$ with feedback phase $\omega_0 \tau = 2\pi m$, $m \in \mathbb{N}$, are presented. The TLS is again excited by a rectangular pulse of duration $\Gamma t_D = 2$ (yellow, shaded area) containing up to three photons. {\color{black} To benchmark the results we obtain with our Heisenberg method (dashed lines) we use the matrix product state framework (solid lines) which is well suited for the simulation of the dynamics in the case of a rectangular pulse shape  before we proceed to more complex pulse shapes as well as an additional dephasing in Sec.~III \cite{Pichler2016,Guimond2017,Barkemeyer2021}.} The agreement of the results obtained with both methods confirms the validity of our approach in the non-Markovian regime.

{\color{black}
\subsection{Comparison with other methods}

There exist different methods for the treatment of WQED setups with coherent time-delayed feedback where it depends on the focus of the investigation, which method is eligible.
Methods such as the space-discretized waveguide model presented in Ref.~\cite{ArranzRegidor2021} treat the light field classically and quantum pulses have not been implemented so far. Other approaches allow for the inclusion of quantum pulses but are restricted to the two-excitation case such as the real-space approach employed in Ref.~\cite{Calajo2019a} or the scattering approach used in Ref.~\cite{Guimond2017} which is particularly suitable if the properties of the light field are of interest.
For treating quantum pulses containing many photons, it is convenient to use the matrix product state framework, which we make use of in Sec.~II to benchmark the Heisenberg method proposed here. 
Compared to the time evolution with matrix product states, however, our Heisenberg method has two major advantages.
First, the inclusion of arbitrary pulse shapes in the matrix product state framework is cumbersome \cite{Barkemeyer2021}. In particular for long pulses, the required decomposition of the initial reservoir state into the MPS form quickly becomes infeasible analytically as well as numerically. Employing the Heisenberg method, in contrast, arbitrary pulse shapes can be included by directly implementing the pulse shape $f(t)$ in the set of differential equations given in Eqs.~\eqref{eq:E_mat} and \eqref{eq:P_mat}.
Furthermore, the straightforward compatibility of the Heisenberg method with additional dissipation channels distinguishes it from other approaches such as the matrix product state framework, where the extension to include non-hermitian dissipation channels requires a substantial numerical overhead \cite{Verstraete2004,Zwolak2004,Daley2009,Kaestle2021}.
Results for corresponding scenarios where the Heisenberg approach is used are presented below in Sec.~III.
In addition to these points, the matrix product state approach is most efficient for small delay times. Large delay times require a multitude of swapping operations, which slow down the algorithm, whereas the influence of the delay time for the Heisenberg approach is negligible.

}

\section{Results}
\label{sec:Results}

The method we introduced above allows the efficient simulation of multiphoton-pulse dynamics for {\color{black} WQED setups under the influence of feedback.} We illustrate the capabilities of the approach by
studying the excitation of an atom-photon bound state for different pulse shapes and system parameters and, furthermore, show that within the Heisenberg formalism a phenomenological pure dephasing rate can be included as an additional dissipation channel.

\subsection{Bound state excitation}

\begin{figure}
    \centering
    \includegraphics[width=8.6cm]{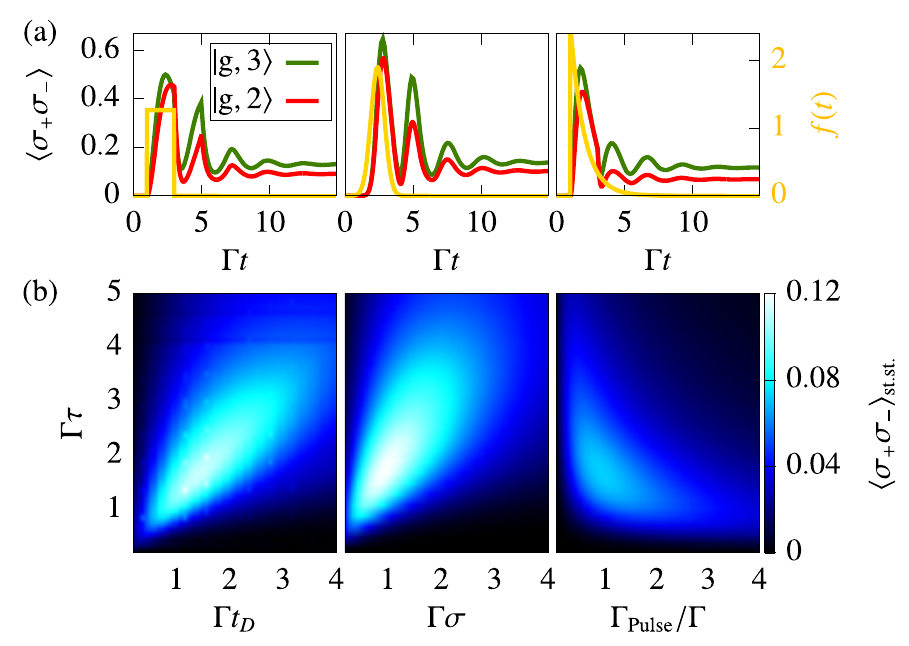}
    \caption{(Color online) {\color{black}Excitation of the atom-photon bound state via multiphoton pulses. (a) Expectation value of the TLS population as a function of time for a TLS with feedback at $\Gamma \tau = 2$ excited via a pulse containing two (red [medium gray] lines) or three photons (green [dark gray] lines)  of shape $f(t)$ (yellow [light gray] lines). 
    From left to right: 
    Rectangular pulse with $\Gamma t_D = 2$, Gaussian pulse with $\Gamma \sigma = 1$, exponentially decaying pulse with $\Gamma_{\text{Pulse}}/\Gamma = 1$.}
    (b)
    Steady-state population of the TLS in the case of a two-photon pulse as a function of the width of the pulse and the feedback delay time, $\tau$.
    From left to right:
    Rectangular pulse with duration $t_D$, Gaussian pulse with width $\sigma$, exponentially decaying pulse with decay rate $\Gamma_\text{Pulse}$.}
    \label{fig:Plot_multi}
\end{figure}

We subject the TLS inside the semi-infinite waveguide to multiphoton pulses of different shapes $f(t)$.
A non-zero population of the emitter in the long-time limit despite the radiative decay is possible if an atom-photon bound state is excited. 
In this case, the emission from the TLS and the feedback signal interfere, and the excitation is partially trapped between emitter and mirror \cite{Gonzalez-Ballestero2013,Tufarelli2013,Redchenko2014,Carmele2020}. 
A prerequisite for the excitation of this bound state in the continuum of propagating modes is a feedback phase $\omega_0 \tau = 2\pi m$, $m \in \mathbb{N}$, which we assume to be realized in our system via a suitable emitter-mirror distance.
Only multiphoton pulses, that is, pulses containing two or more photons, can excite the atom-photon bound state while an emitter subjected to a single-photon pulse necessarily decays to the ground state in the long-time limit \cite{Calajo2019a}. 
It is the intrinsic nonlinearity of the TLS that allows the excitation of the bound state via stimulated emission which, however, only comes into play if the emitter is excited with at least two photons.
Another possibility to address the atom-photon bound state is by letting an initially excited emitter decay in vacuum \cite{Tufarelli2013}. While for this scheme, the steady-state population decreases monotonously with the delay time $\tau$, the behavior is more complex in the case of multiphoton pulses.

{\color{black}
In Fig.~\ref{fig:Plot_multi}, the excitation of an atom-photon bound state is studied for three different pulse shapes: We consider a rectangular pulse with $f(t) = A \Theta(t-t_0)\Theta(t_0+t_D-t)$ where $A$ is a constant ensuring that the normalization condition $\int dt \left| f(t)\right|^2 = 1$ is fulfilled, $t_0$ is the starting point, and $t_D$ is the duration of the pulse. Furthermore, a Gaussian pulse is applied with $f(t)= A \exp\left[-(t-\mu)^2/(2\sigma^2) \right]$ where $A$ again is a normalization constant, $\mu$ is the offset of the pulse, and $\sigma$ determines its width. Finally, we subject the system to an exponentially decaying pulse characterized by $f(t) = A \Theta(t-t_0)\exp\left[-\Gamma_\text{Pulse}(t-t_0)\vphantom{a^2}\right]$ with normalization constant $A$, starting point $t_0$, and decay constant $\Gamma_\text{Pulse}$. 
In Fig.~\ref{fig:Plot_multi}\,a, the dynamics of the expectation value of the TLS population in the case of pulses containing two (red [medium gray] lines) or three photons (green [dark gray] lines) are shown. The TLS is subjected to feedback at $\Gamma \tau = 2$ and excited via pulses of shape $f(t)$ (yellow [light gray] lines) where the considered pulses are a rectangular pulse with $\Gamma t_D = 2$, a Gaussian pulse with $\Gamma \sigma = 1$, and an exponentially decaying pulse with $\Gamma_{\text{Pulse}}/\Gamma = 1$ (from left to right).
For all considered pulse shapes, the TLS is excited and, after a transient time, stabilizes at a non-trivial steady-state population which indicates the excitation of the atom-photon bound state. Furthermore, for the chosen parameters, increasing the number of photons in the pulses results in an increased steady-state excitation pointing to an enhanced excitation efficiency of the bound state.

In Fig.~\ref{fig:Plot_multi}\,b, the steady-state population of the emitter as a function of the pulse width and the feedback delay time is shown for the three different fundamental pulse shapes containing two photons, $\braket{\sigma_+\sigma_-}_\text{st.st.} \equiv \lim_{t \rightarrow \infty}\braket{g,2|\sigma_+(t)\sigma_-(t)|g,2}$. 
Our method is ideally suited for such a parameter study since it allows the exact and efficient calculation of the dynamics as well as the straightforward inclusion of arbitrary pulse shapes.
It becomes apparent that the interplay of the feedback time and the pulse width plays a crucial role for the steady-state population the emitter is stabilized at and we see that a certain minimum separation between emitter and mirror is crucial for the possibility to excite the bound state by multiphoton scattering. Furthermore, we observe that, in the case of a rectangular pulse shape, the highest steady-state population is found for $\tau \approx t_\text{D}$, so that the feedback sets in approximately at the moment the pulse has ended.%
}

\begin{figure}
    \centering
    \includegraphics[width=8.6cm]{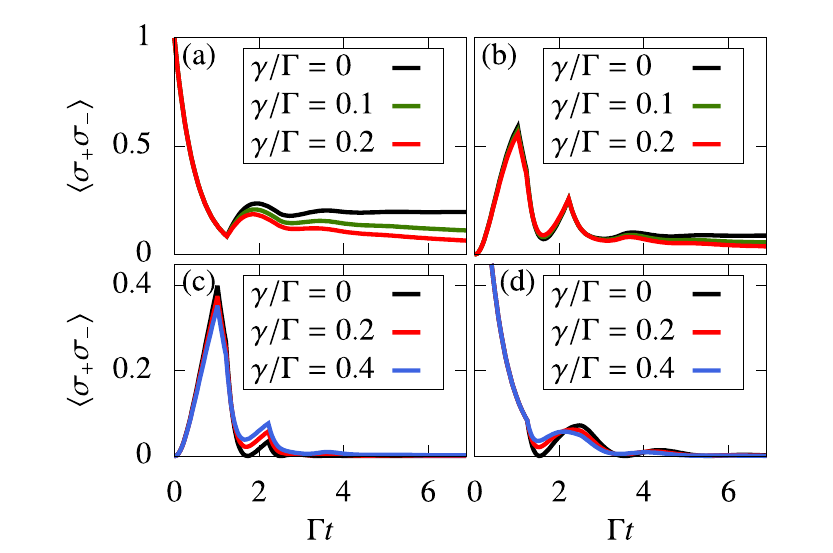}
    \caption{(Color online) Expectation value of the TLS population as a function of time under the influence of pure dephasing at different rates $\gamma$ for feedback at $\Gamma \tau=1.2$ for a system in the initial state $\ket{\psi}$. (a) Initially excited emitter decaying in vacuum ($\ket{\psi} = \ket{e,0}$) with feedback phase $\omega_0 \tau = 2\pi m$, $m \in \mathbb{N}$. (b) Emitter initially in the ground state excited via a rectangular two-photon pulse of duration $\Gamma t_D = 1$ ($\ket{\psi} = \ket{g,2}$) with feedback phase $\omega_0 \tau = 2\pi m$. (c) Emitter initially in the ground state excited via a rectangular single-photon pulse of duration $\Gamma t_D = 1$ ($\ket{\psi} = \ket{g,1}$) with feedback phase $\omega_0 \tau = 2\pi m$. (d) Initially excited emitter decaying in vacuum ($\ket{\psi} = \ket{e,0}$) with feedback phase $\omega_0 \tau = (2 m + 1)\pi$.}
    \label{fig:PD}
\end{figure}

\subsection{Additional dissipation channel}
In realistic WQED systems, a complete isolation of the system of interest from its environment is not possible and typically leads to energy dissipation and loss of quantum coherence. Thus, the dynamics are effectively rendered non-unitary. If the setup is, for example, realized in a solid-state environment where semiconductor quantum dots function as the emitters, the interaction with the surrounding bulk material has to be taken into consideration \cite{Krummheuer2002,Ramsay2010,Kuhlmann2013,Roy-Choudhury2015,Nazir2016,Iles-Smith2017,Reiter2019}.

As a first proof of principle, we include a phenomenological pure dephasing at rate $\gamma$ and study its impact on the TLS population. Pure dephasing affects coherences while it does not have an immediate influence on populations. Therefore, to model pure dephasing, we treat the coherence operator $\sigma_-$ and the TLS population operator $E \equiv \sigma_+\sigma_-$ separately and introduce pure dephasing in the differential equations for the matrix elements of the coherence operator as an additional Markovian decay channel, see Appendix~C. In this model, for example, the matrix element $\bra{g,0}\sigma_-(t) \ket{g,1}$ which is relevant in the case of a single-photon pulse in the reservoir initially, obeys the differential equation
\begin{multline}
    \frac{\text{d}}{\text{d}t} \bra{g,0}\sigma_-(t) \ket{g,1} = -(\Gamma+\gamma) \bra{g,0}\sigma_-(t) \ket{g,1} - \sqrt{\Gamma}f_\tau(t) \\
    + \Gamma e^{i\omega_0 \tau} \bra{g,0}\sigma_-(t-\tau) \ket{g,1} \Theta(t-\tau).
\end{multline}

The results for the dynamics of the expectation value of the TLS population for different pure dephasing rates $\gamma$ are shown in Fig.~\ref{fig:PD}. In the setup, feedback at the delay time $\Gamma \tau = 1.2$ is implemented and different scenarios are considered. 
For an initially excited emitter decaying in vacuum, as illustrated in Fig.~\ref{fig:PD}\,a, pure dephasing does not influence the TLS population before the feedback mechanism sets in. Since it only dephases the coherence, its effect emerges for $t > \tau$ and renders the long-time stabilization of the excitation impossible. In the considered setup, a feedback phase of $\omega_0 \tau = 2\pi m$, $m \in \mathbb{N}$, is assumed as a prerequisite for the stabilization.
For a dephasing rate $\gamma \neq 0$, however, we find a decay of the TLS population where it holds that the higher the dephasing rate, the faster the decay. 

In the next step, we assume the TLS to be initially in the ground state and excited via a two-photon pulse. The dynamics that can be observed in the case of a rectangular pulse of duration $\Gamma t_D = 1.2$ are shown in Fig.~\ref{fig:PD}\,b. In this case, the dephasing already influences the dynamics before the first feedback round trip is completed and reduces the effectiveness of the excitation of the emitter. 
The higher the dephasing rate compared to the pulse width, the less effective the excitation, since the dephasing
prevents the build-up of coherence. In addition, the atom-photon bound state becomes inaccessible with pure dephasing and the emitter inevitably decays to the ground state.
In Fig.~\ref{fig:PD}\,c, an emitter subjected to a single-photon pulse is considered. Here, a dephasing at higher rates than considered for Figs.~\ref{fig:PD}\,a and b results in an even more apparent reduction of the excitation effectiveness. Furthermore, we see that the dephasing is not necessarily detrimental at all times. Since it counteracts constructive as well as destructive interference, dephasing can result in a population that temporarily exceeds the one observed in the case without dephasing.

After concentrating on the case of a feedback phase $\omega_0 \tau = 2\pi m$ so far, we now look at the impact of pure dephasing for an initially excited TLS decaying in vacuum with feedback at phase $\omega_0\tau = (2m+1)\pi$ as illustrated in Fig.~\ref{fig:PD}\,d. At this feedback phase, without dephasing, the decay of the TLS population is maximally accelerated. With dephasing, we observe that the effect of the feedback is diminished and the dynamics are approaching the exponential Wigner-Weißkopf decay we observe in the absence of feedback. 

\section{Conclusion and outlook}
We presented a Heisenberg-operator approach for efficiently calculating the non-Markovian dynamics in WQED systems with feedback. The hierarchical structure of multi-time correlations which typically arises when treating such systems in the Heisenberg picture can be unraveled via the insertion of a unity operator between the operators with different time arguments. This way, the complexity of the problem is shifted to the calculation of matrix elements of single-time Heisenberg operators. We introduced the method using the example of a TLS inside a semi-infinite waveguide which can be excited via multiphoton pulses. For this setup, we demonstrated that our method is a versatile tool that allows the consideration of arbitrary pulse shapes as well as the inclusion of additional dissipation channels.

We focussed particularly on the atom-photon bound state that exists in the system, and studied the complex interplay of the feedback delay time and the pulse shape which determine the excitation efficiency of the bound state. Furthermore, we showed that the bound state is inaccessible in the presence of Markovian pure dephasing. 

In future research, we plan to use the proposed method to study the impact of interactions in the many-body context in more detail. {\color{black}To that end, it could be interesting to combine the proposed method with other approaches such as matrix product state or quantum trajectory simulations and make use of the advantages of each approach.}

\begin{acknowledgements}
The authors gratefully acknowledge the support of the Deutsche Forschungsgemeinschaft (DFG) through the project B1 of the SFB 910 (Project No. 163436311). The authors furthermore thank Oliver Kästle for fruitful discussions.
\end{acknowledgements}

\begin{appendices}
\section*{Appendix A: Dynamics without feedback}

\setcounter{equation}{0}
\renewcommand{\theequation}{A\arabic{equation}}

Before the feedback sets in, that is, for times $t<\tau$, the dynamics are essentially Markovian and no multi-time correlations arise. The expectation value of the TLS population for $n$ photons in the reservoir initially, $\bra{g,n}\sigma_+(t)\sigma_-(t) \ket{g,n}$, can be calculated from Eq.~\eqref{eq:P_op} of the main text recursively using the results for up to $n-1$ photons in the reservoir. In this case, the dynamics are governed by 
\begin{align}
    \frac{\text{d}}{\text{d}t}&\bra{g,n}\sigma_+(t)\sigma_-(t)\ket{g,n} = -2\Gamma\bra{g,n}\sigma_+(t)\sigma_-(t)\ket{g,n} \notag \\
    &- \sqrt{n \Gamma}\left[ f_\tau^*(t) \bra{g,n-1} \sigma_-(t) \ket{g,n} + \text{H.c.}\right], \\
    \frac{\text{d}}{\text{d}t}&\bra{g,n-1}\sigma_-(t)\ket{g,n} = -\Gamma \bra{g,n-1}\sigma_-(t)\ket{g,n} \notag \\
    &- \sqrt{n\Gamma}f_\tau(t) \left[1-2\bra{g,n-1}\sigma_+(t)\sigma_-(t) \ket{g,n-1} \right].
\end{align}

\section*{Appendix B: Dynamics under the influence of feedback}

\setcounter{equation}{0}
\renewcommand{\theequation}{B\arabic{equation}}

For larger times $t \geq \tau$ where the feedback mechanism influences the dynamics, we insert the unity operator given in Eq.~\eqref{eq:unity} of the main text so that we only have to deal with single-time matrix elements of the TLS operator $\sigma_-$.
Here, we present the calculations for one and two photons in the reservoir initially. The extension to three and more photons then works analogously.

\subsection*{Single-photon pulse}

For an initial state $\ket{g,1}$, only the projector onto the state $\ket{g,0}$ describing the TLS in the ground state and no photons in the reservoir contributes when decomposing the expectation value into matrix elements of the TLS operator $\sigma_-$ and it holds that
\begin{multline}
    \bra{g,1}\sigma_+(t)\sigma_-(t)\ket{g,1} = \\ \bra{g,1}\sigma_+(t)\ket{g,0}\bra{g,0}\sigma_-(t)\ket{g,1}.
\end{multline}
Thus, the only relevant matrix element is $\bra{g,0}\sigma_-(t)\ket{g,1}$ which can be determined via the differential equation
\begin{multline}
   \frac{\text{d}}{\text{d}t} \bra{g,0}\sigma_-(t)\ket{g,1} = -\Gamma \bra{g,0}\sigma_-(t)\ket{g,1} \\
   - \sqrt{\Gamma}f_\tau(t)+\Gamma e^{i\omega_0\tau} \bra{g,0}\sigma_-(t-\tau)\ket{g,1} \Theta(t-\tau).
\end{multline}

\subsection*{Two-photon pulse}

In the case of an initial state $\ket{g,2}$, the projector onto states describing one excitation in the system, either in the TLS or in the reservoir, contributes to the dynamics and we can decompose the expectation value of the TLS population according to
\begin{multline}
\bra{g,2}\sigma_+(t)\sigma_-(t)\ket{g,2} =  \bra{g,2}\sigma_+(t)\ket{e,0}\bra{e,0}\sigma_-(t)\ket{g,2}  \\
+ \int dt'\bra{g,2}\sigma_+(t)\ket{g,t'}\bra{g,t'}\sigma_-(t)\ket{g,2}.
\end{multline}
The matrix elements of the TLS operator $\sigma_-$ describing the transition from two to one excitation in the system that have to be evaluated evolve as
\begin{widetext}
\begin{multline}
    \frac{\text{d}}{\text{d}t} \bra{e,0}\sigma_-(t)\ket{g,2}
    = -\Gamma \bra{e,0}\sigma_-(t)\ket{g,2} +2 \sqrt{2\Gamma}f_\tau(t) \bra{e,0}\sigma_+(t)\ket{g,0}\bra{g,0}\sigma_-(t)\ket{g,1}
    +\Gamma e^{i\omega_0\tau} \left\{ \bra{e,0}\sigma_-(t-\tau)\ket{g,2} \vphantom{\int} \right. \\
    \left. - 2 \left[ \bra{e,0}\sigma_+(t)\ket{g,0}\bra{g,0}\sigma_-(t)\ket{e,0}\bra{e,0}\sigma_-(t-\tau)\ket{g,2} + \int dt_1 \bra{e,0}\sigma_+(t)\ket{g,0}\bra{g,0}\sigma_-(t)\ket{g,t_1}\bra{g,t_1}\sigma_-(t-\tau)\ket{g,2}  \right]\right\} \\ \times \Theta(t-\tau)
\end{multline}
and
\begin{multline}
    \frac{\text{d}}{\text{d}t} \bra{g,t'}\sigma_-(t)\ket{g,2} = -\Gamma \bra{g,t'}\sigma_-(t)\ket{g,2} - \sqrt{2\Gamma}f_\tau(t)\left[ \braket{g,t'|g,1}-2 \bra{g,t'}\sigma_+(t)\ket{g,0}\bra{g,0}\sigma_-(t)\ket{g,1}\right] \\
    +\Gamma e^{i\omega_0\tau}\left\{ \bra{g,t'}\sigma_-(t-\tau)\ket{g,2} - 2 \left[ \bra{g,t'} \sigma_+(t)\ket{g,0}\bra{g,0}\sigma_-(t)\ket{e,0}\bra{e,0}\sigma_-(t-\tau)\ket{g,2}\vphantom{\int} \right.\right. \\ \left.\left. + \int dt_1 \bra{g,t'}\sigma_+(t)\ket{g,0}\bra{g,0}\sigma_-(t)\ket{g,t_1}\bra{g,t_1}\sigma_-(t-\tau)\ket{g,2} \right]\right\} \Theta(t-\tau)
\end{multline}
\end{widetext}
with $ \braket{g,t'|g,1} = f(t')$.
These equations, in turn, couple to matrix elements of $\sigma_-$ that refer to the transition from one to zero excitations. The matrix elements that contribute in addition to the element $\bra{g,0}\sigma_-(t)\ket{g,1}$ which has already been introduced in the single-excitation case above obey
\begin{multline}
     \frac{\text{d}}{\text{d}t} \bra{g,0}\sigma_-(t)\ket{e,0} = -\Gamma \bra{g,0}\sigma_-(t)\ket{e,0}
    \\
    +\Gamma e^{i\omega_0\tau} \bra{g,0}\sigma_-(t-\tau)\ket{e,0}  \Theta(t-\tau)
\end{multline}
and
\begin{multline}
     \frac{\text{d}}{\text{d}t}\bra{g,0}\sigma_-(t)\ket{g,t'} = -\Gamma \bra{g,0}\sigma_-(t)\ket{g,t'}
     \\- \sqrt{\Gamma} \bra{g,0}r_{t,\tau}\ket{g,t'}+\Gamma e^{i\omega_0\tau} \bra{g,0}\sigma_-(t-\tau)\ket{g,t'} \\  \times \Theta(t-\tau)
\end{multline}
with
$ r_{t,\tau}\ket{g,t'} = \left[\delta\left(t'-t+\frac{\tau}{2}\right)e^{i\omega_0\frac{\tau}{2}}-\delta\left(t'-t-\frac{\tau}{2}\right)e^{-i\omega_0\frac{\tau}{2}}\right]\ket{g,0}$.

\section*{Appendix C: Implementation of pure dephasing}

\setcounter{equation}{0}
\renewcommand{\theequation}{C\arabic{equation}}

Pure dephasing dephases coherences while populations are not directly affected. To be able to account for this fact, not all matrix elements are decomposed into single-time matrix elements of the operator $\sigma_-$ but are distinguished into the coherence operator $\sigma_-$ and the TLS population operator $E\equiv \sigma_+\sigma_-$. 
This way, for the calculation of the dynamics for a TLS initially in the ground state and a single-photon pulse in the reservoir, we have to solve
\begin{align}
    &\frac{\text{d}}{\text{d}t} \bra{g,1}E(t)\ket{g,1} =  -2\Gamma \bra{g,1}E(t)\ket{g,1} \notag \\
    &- \sqrt{\Gamma}\left[f_\tau^*(t)\bra{g,0}\sigma_-(t)\ket{g,1} + \text{H.c.}\right]\notag \\
    &+\Gamma\left[e^{-i\omega_0\tau}\bra{g,1}\sigma_+(t-\tau)\sigma_-(t)\ket{g,1} + \text{H.c.}\right]  \Theta(t-\tau).
\end{align}
The insertion of a unity between the operators with different time arguments yields
\begin{align}
    &\frac{\text{d}}{\text{d}t} \bra{g,1}E(t)\ket{g,1} =  -2\Gamma \bra{g,1}E(t)\ket{g,1} \notag \\
    &- \sqrt{\Gamma}\left[f_\tau^*(t)\bra{g,0}\sigma_-(t)\ket{g,1} + \text{H.c.}\right]\notag \\
    &+\Gamma\left[e^{-i\omega_0\tau}\bra{g,1}\sigma_+(t-\tau)\ket{g,0}\bra{g,0}\sigma_-(t)\ket{g,1} + \text{H.c.}\right] \notag \\ 
    &\qquad \qquad \qquad \qquad \qquad \qquad \qquad \qquad \quad \times\Theta(t-\tau).
\end{align}
The expectation value couples to the matrix element $\bra{g,0}\sigma_-(t)\ket{g,1}$ for which we add a phenomenological pure dephasing at rate $\gamma$ so that it obeys
\begin{multline}
    \frac{\text{d}}{\text{d}t} \bra{g,0}\sigma_-(t) \ket{g,1} = -(\Gamma+\gamma) \bra{g,0}\sigma_-(t) \ket{g,1} - \sqrt{\Gamma}f_\tau(t) \\
    + \Gamma e^{i\omega_0 \tau} \bra{g,0}\sigma_-(t-\tau) \ket{g,1} \Theta(t-\tau).
\end{multline}

Analogously, for a two-photon pulse in the reservoir, we have to evaluate the expectation value
\begin{align}
    &\frac{\text{d}}{\text{d}t} \bra{g,2}E(t)\ket{g,2} = -2\Gamma \bra{g,2}E(t)\ket{g,2} \notag\\
    &- \sqrt{2\Gamma}\left[f_\tau^*(t)\bra{g,1}\sigma_-(t)\ket{g,2} + \text{H.c.}\right] \notag \\
    &+\Gamma\left\{e^{-i\omega_0\tau}\left[\bra{g,2}\sigma_+(t-\tau)\ket{e,0}\bra{e,0}\sigma_-(t)\ket{g,2} \vphantom{\int} \right.\right. \notag \\ &\left.\left. + \int d t'\bra{g,2}\sigma_+(t-\tau)\ket{g,t'}\bra{g,t'}\sigma_-(t)\ket{g,2}\right]  + \text{H.c.}\right\} \notag \\
    &\qquad \qquad \qquad \qquad \qquad \qquad \qquad \qquad \quad \times\Theta(t-\tau).
\end{align}
After the introduction of a phenomenological pure dephasing at rate $\gamma$, the matrix element $\bra{g,1}\sigma_-(t)\ket{g,2}$ which couples to the dynamics obeys the differential equation 
\begin{align}
    &\frac{\text{d}}{\text{d}t} \bra{g,1}\sigma_-(t) \ket{g,2} = - (\Gamma +\gamma) \bra{g,1}\sigma_-(t) \ket{g,2} \notag \\
    & \qquad - \sqrt{2\Gamma}f_\tau(t) \left[1-2\bra{g,1}E(t)\ket{g,1} \right] \notag \\
     &+ \Gamma e^{i\omega_0\tau}\left[\vphantom{\int}\bra{g,1}\sigma_-(t-\tau)\ket{g,2} \right. \notag \\ &\left.- 2 \bra{g,1}E(t)\ket{e,0}\bra{e,0}\sigma_-\ket{g,2} \right. \notag \\ &\left. - 2 \int\mathrm{d}t' \bra{g,1}E(t)\ket{g,t'}\bra{g,t'}\sigma_-(t-\tau)\ket{g,2} \right]\notag \\
    &\qquad \qquad \qquad \qquad \qquad \qquad \qquad \qquad \quad \times\Theta(t-\tau).
\end{align}
The remaining elements that contribute to the dynamics can be treated analogously until, eventually, the system of differential equations closes. 

The calculation of the dynamics for other initial states such as $\ket{e,0}$ describing the TLS initially in the excited state and the reservoir in the vacuum state works analogously.
For the differentiation into matrix elements of the operators $\sigma_-$ and $E$ the number of matrix elements that have to be determined and saved increases. The asymptotic runtime, however, remains unchanged.
\end{appendices}

\bibliographystyle{apsrev4-2}
\bibliography{Bibfile}

\end{document}